\documentclass[apj]{emulateapj}
\newcommand{\myemail}{Lauren.MacArthur@nrc-cnrc.gc.ca}
\newcommand{\alphaFe}{[$\alpha$/Fe]}
\newcommand{\etal}{et~al.\@}
\newcommand{\eg}{e.g.\@,}
\newcommand{\ie}{i.e.\@,}
\newcommand{\hii}{\ion{H}{2}}
\newcommand{\chisqr}{$\chi^2$}
\newcommand{\taueff}{$\hat{\tau}$}
\newcommand{\avgAl}{\ifmmode {\langle{\rm A}\rangle_{l}} \else {$\langle$A$\rangle_{l}$} \fi}
\newcommand{\avgZl}{\ifmmode {\langle{\rm Z}\rangle_{l}} \else {$\langle$Z$\rangle_{l}$} \fi}

\shorttitle{Integrated Stellar Populations: Photometry versus Spectroscopy}

\begin{document}

\title{INTEGRATED STELLAR POPULATIONS: CONFRONTING PHOTOMETRY WITH SPECTROSCOPY}


\author {Lauren A. MacArthur\altaffilmark{1,2},
Michael McDonald\altaffilmark{3}, 
St{\' e}phane Courteau\altaffilmark{4}, and 
J. Jes{\' u}s Gonz{\'a}lez\altaffilmark{5}}

\altaffiltext{1}{Herzberg Institute of Astrophysics, National Research 
                Council of Canada, 5071 West Saanich Road, Victoria, BC 
                V8X 4M6, Canada; \myemail}
\altaffiltext{2}{Department of Physics \& Astronomy, University of Victoria,
	         Victoria, BC, V8P 1A1, Canada}
\altaffiltext{3}{Department of Astronomy, University of Maryland,
                 College Park, MD 20742-2421, USA}
\altaffiltext{4}{Department of Physics, Engineering Physics \& Astronomy, 
                 Queen's University, Kingston, ON K7L 3N6, Canada}
\altaffiltext{5}{Instituto de Astronomia, Universidad Nacional Aut{\'o}noma 
                de M{\'e}xico, Apdo Postal 70-264, Cd. Universitaria, 
                04510, Mexico}

%\date{\today}

\begin{abstract}
We investigate the ability of spectroscopic techniques to yield
realistic star formation histories (SFHs) for the bulges of spiral
galaxies based on a comparison with their observed broadband colors.
Full spectrum fitting to optical spectra indicates that recent (within
$\sim$\,1\,Gyr) star formation activity can contribute significantly
to the $V$-band flux, whilst accounting for only a minor fraction of
the stellar mass budget which is made up primarily of old stars.
Furthermore, recent implementations of stellar population (SP) models
reveal that the inclusion of a more complete treatment of the thermally
pulsating asymptotic giant branch (TP-AGB) phase to SP models greatly
increases the NIR flux for SPs of ages 0.2--2\,Gyr.  Comparing the
optical--NIR colors predicted from population synthesis fitting, using
models which do not include all stages of the TP-AGB phase, to the
observed colors reveals that observed optical--NIR colors are too red
compared to the model predictions.  However, when a 1\,Gyr SP from
models including a full treatment the TP-AGB phase is used, the
observed and predicted colors are in good agreement.  This has strong
implications for the interpretation of stellar populations, dust
content, and SFHs derived from colors alone.
\end{abstract}


\section{Introduction}\label{sec:intro}

A detailed understanding of the stellar populations (SPs) that make up
the integrated spectral energy distributions (SEDs) of both local and
distant galaxies can provide important constraints for models of
galaxy formation.  However, confronting observations with SP
synthesis models for the purpose of translating the former into
physical parameters, is not without significant challenges.  While
SP modeling has seen tremendous progress over the past decade (\eg\
Bruzual \& Charlot 2003, hereafter BC03; Le~Borgne \etal\ 2004;
Maraston 2005, hereafter Mar05; Schiavon 2007), misinterpretations in
the data--model comparison are still common.

Broadband colors are often used as a proxy for SP parameters.  The
well-known age/$Z$ degeneracy in optical colors is greatly lifted by
the addition of NIR bands.  However, in broadband-based analyses,
extinction and reddening effects from interstellar dust cannot be
ruled out.  Indeed, observed optical--NIR colors that lie redward of
the model grids, \ie\ in a region not supported by a naked SP of any
age/$Z$/SFH, have typically been attributed to dust reddening (\eg\
Peletier \etal\ 1999; MacArthur
\etal\ 2004; Carollo \etal\ 2007).  Meanwhile, red optical--NIR colors 
within the model grids are assumed to result from old and metal-rich
SPs.  

In light of recent implementations of SP models which account for a
more detailed treatment of the effects of the thermally pulsating
asymptotic giant branch (TP-AGB) stellar evolutionary phase (\eg\
Mar05; Coelho \etal\ 2007; Lee \etal\ 2009), a reassessment of the
observed red optical--NIR colors in galaxies is in order.  Accounting
for the TP-AGB is accomplished in Mar05 through empirical calibration to
LMC globular clusters which have independent age and $Z$ measurements.
Alternatively, synthetic model tracks can be used.  The recent tracks
of Marigo \& Girardi (2007), whose predictions are in good agreement
with those of Mar05, account for nine evolutionary stages (including the
C-, M-type, and superwind mass loss phases) in the TP-AGB.  (Note that
the older tracks used in the BC03 models account for only a single
evolutionary stage for each evolutionary phase in the TP-AGB.)  In
these models, the TP-AGB phase is active in SPs of ages
$\sim$\,0.3--2\,Gyr and contributes significantly to the NIR flux
(accounting for up to $\sim$\,80\% of the NIR flux), leading to very
red optical--NIR colors for this age range.  These models have already
been used to resolve the uncomfortably large stellar masses and old
ages derived for high-$z$ galaxies using models which do not account
for all phases in their treatment of the TP-AGB phase (Maraston \etal\
2006).

In the meantime, full spectrum fitting techniques have proven to be
effective at recovering the underlying stellar content of
integrated galaxy spectra (\eg\ Heavens \etal\ 2000; Cid~Fernandes
\etal\ 2005; Walcher \etal\ 2006; MacArthur, Gonz{\' a}lez, \& 
Courteau 2009, hereafter Mac09).  This type of analysis provides a
stochastically-sampled SFH for a given integrated spectrum and allows
for true average, as opposed to single SP-based, age and $Z$
estimates.  Full spectral fitting is typically limited to optical
wavelengths, as set by the observations.  However, as most models
provide SEDs in the UV--FIR range, one can use such model fits to make
predictions for the galaxy light in other wavelength regions.  For
example, a comparison of predicted versus observed optical--NIR colors
could provide valuable insight into both the reliability of the
spectral fits as well as any shortcomings of the SP models themselves.
This is precisely the approach we carry out here using our analysis of
Gemini/GMOS long-slit spectra of local spirals (Mac09) and new optical
imaging for the same galaxies from the Palomar Observatory and NIR
imaging from the Two Micron All Sky Survey (hereafter 2MASS; 
Skrutskie \etal\ 2006).  As we are interested in predictions from models
that use different treatments of the TP-AGB phase, we focus on two
galaxies from our spectroscopic sample (NGC 628
\& UGC 2124) with the highest quality data and whose derived SFHs 
indicate significant contribution (in $V$-band light-weight) from a
1\,Gyr SP, \ie\ where the TP-AGB NIR signature is expected to be
strong.

\section{Data}\label{sec:data}
The long-slit spectroscopic data used for this study, from Mac09,
were collected with the Gemini Multi-Object Spectrograph (GMOS; Hook
\etal\ 2004) on the 8-m Gemini North telescope.
The GMOS detector and B600 grating combination provides a
spatial resolution of 0.072\arcsec/pix and a dispersion of
0.45\,\AA/pix. The slit field of view (FOV) of
5\arcmin\,$\times$\,2\arcsec\ provided a boxcar width\,=\,10.81\,\AA\
resolution.  The spectral coverage spans $\sim$\,4050\,--\,6750\,\AA.

Optical imaging in the $BVRI$ Bessel filters was collected at the Palomar
200-inch telescope in 2006 December. The Large Format Camera consists
of a six CCD mosaic with a FOV of 6$\times$12\arcmin\ per CCD.  Our
galaxies have optical diameters $\le$\,10.5\arcmin, leaving enough
room beyond the optical radius on a single chip for sky measurements.
Typical exposure times were 1--2 minutes and 2$\times$2 binning gave
0.363\arcsec/pixel. The seeing ranged from FWHM\,=\,1.8--3.7\arcsec.
The data were bias-subtracted and flat-fielded using a combination of
night-sky and dome flats to carefully account for both the
high-frequency spatial sensitivity (dome) and the large-scale
illumination (sky) of the chip.  Due to the bright-moon conditions, it
was necessary to model the background with a surface. This induced a
typical sky error of $\sim$\,0.01\,mag. The data were calibrated
following the technique of Courteau (1996) using stars from several
Landolt fields (Landolt 1992), for typical calibration errors of
$\sim$\,0.03\,mag.

Imaging in the NIR $JHKs$ filters were obtained from the 2MASS online
database.  The background levels were carefully re-measured for each
image and the photometric zeropoints were taken from the 2MASS
headers.

All data (imaging and spectroscopic) were corrected for Galactic
foreground extinction (Schlegel, Finkbeiner, \& Davis 1998).

\subsection{Radial Extractions}\label{sec:extract}
In order to make a direct comparison between information gleaned from
optical spectra versus broadband imaging, we must compare the spectral
fits with colors from the same effective radial bins.  As such, rather
than the usual azimuthal extraction of SB profiles (\eg, Courteau
1996; MacArthur, Courteau, \& Holtzman 2003), profiles were extracted
from all bands ($BVRIJHKs$) using the same position angle and slit
width as for the GMOS observations.  We accounted for differences in
the observed point spread functions by convolving all profiles
(imaging and spectroscopic) with the largest FWHM of all observations
for each galaxy.  Finally, radial binning for both the imaging and
spectroscopy was set to the largest measured dispersion, which is that
of 2MASS (1\arcsec/pixel).

\section{Age, Metallicity, and SFH from Population Synthesis}
\begin{figure*}
\begin{center}
\includegraphics[width=0.47\textwidth,bb=18 144 592 518]{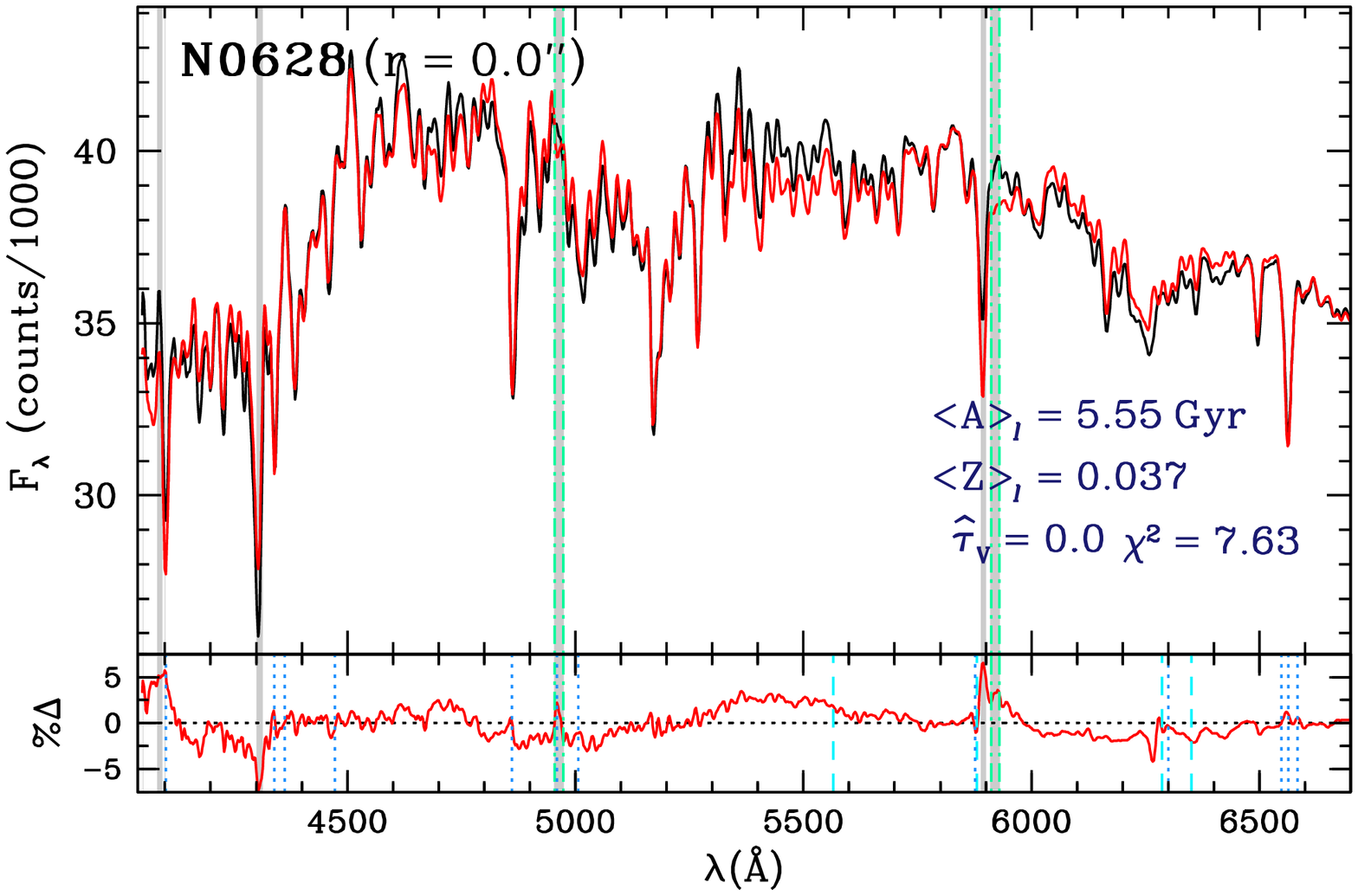} 
\includegraphics[width=0.47\textwidth,bb=18 144 592 518]{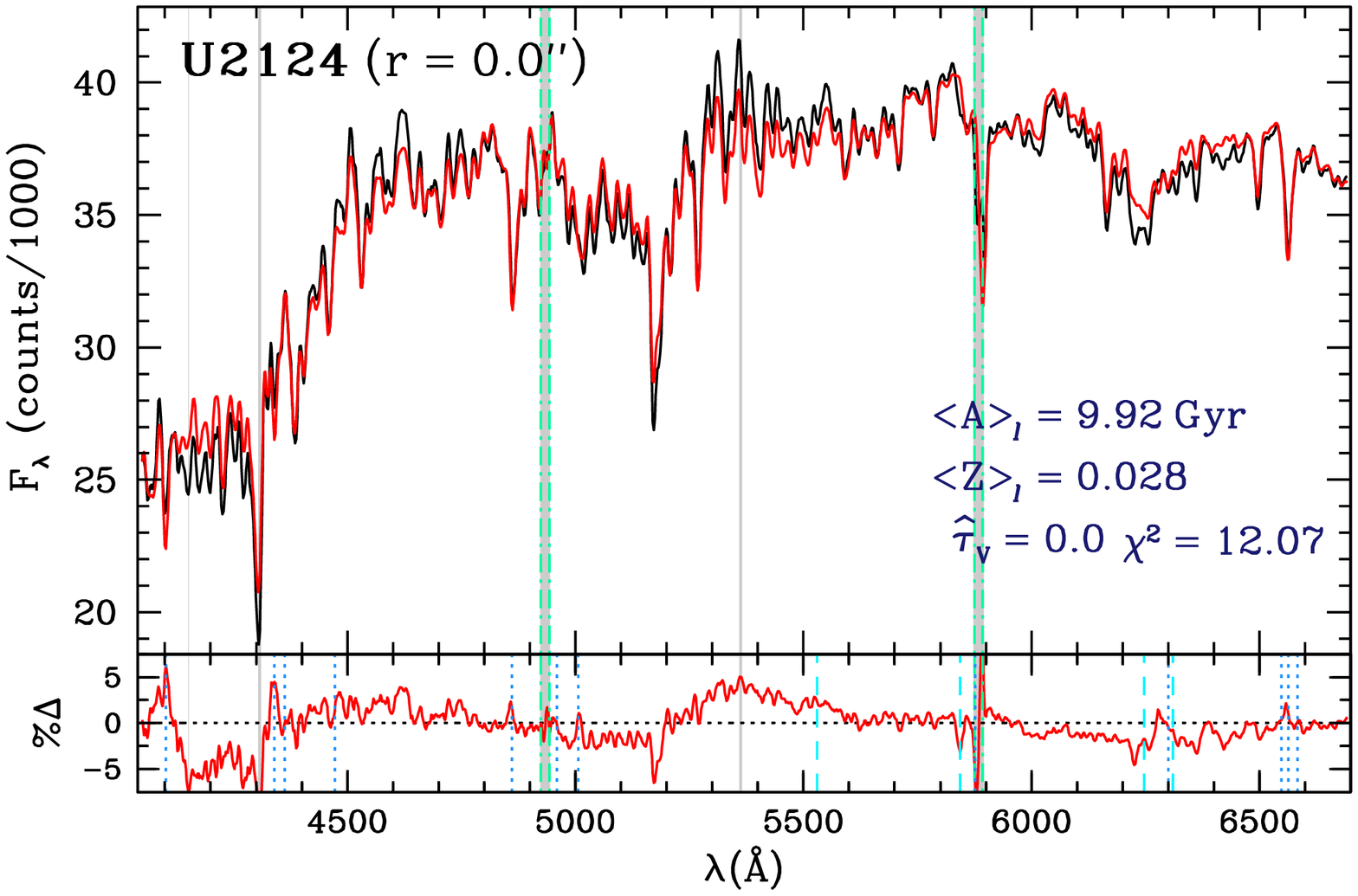}
   \caption{Central observed spectra (black) and full population
   synthesis fit (red) for NGC 628 (left) and UGC 2124 (right).
   Gray shading indicates regions masked in the fit as
   determined by our iterative ``$\sigma$-clipping'' procedure 
   as well as the CCD gap regions (green
   vertical dash-dotted lines) which are always masked.  Shown at
   lower right on each panel are the average light-weighted age,
   \avgAl, metallicity, \avgZl, effective \taueff$_V$, and
   \chisqr\ of the fit.  The bottom panels show the percent
   data$-$model residuals. In the bottom panels, dashed vertical lines
   indicate variable sky-lines, and dotted vertical lines indicate
   emission lines prevalent in star forming (\hii) regions. }
   \label{fig:fits}
\end{center}
\end{figure*}
The SP fitting technique used here is as described in Mac09; a brief
summary is provided below.  Our ``full population synthesis''
technique consists of a bound constrained optimized fit representing
the relative contribution of each of the 70 model templates to the
observed spectrum.  The only fixed bound is that of no negative
template contributions.  The templates are from the BC03 models which
provide SEDs representing simple stellar populations (SSPs), \ie\
single bursts of star formation (SF) at a given age and $Z$, at a
resolution of $\sim$\,3\,\AA\ FWHM\footnote{Although see
\S3.4.1 of Mac09 for a detailed analysis of the true resolution of
these models.}.  The selected library of 70 SSP templates covers the
age range 0.001--20\,Gyr and metallicities of $Z$\,=\,0.0004--0.05 and
we use the models with the Chabrier (2003) initial mass function.  Our
procedure also allows for dust reddening to the observed SED according
to the prescription of Charlot \& Fall (2000). See Mac09 for details.
To accommodate any non-stellar contributions to the observed spectrum
not considered in the BC03 models, we used an iterative masking scheme
whereby deviant $|$data--model$|$ points are given zero weight in
the fit.

\begin{figure}
\begin{center}
\includegraphics[width=0.49\textwidth]{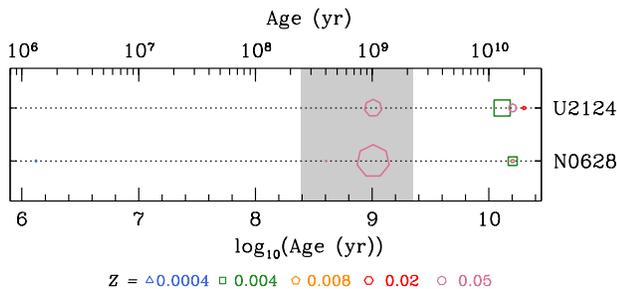}
%\plotone{f2.eps}
\caption{Light-weighted SFHs of the central spectra for NGC 628 and 
         UGC 2124 derived from the full population synthesis fits.
         The dotted horizontal lines guide the eye for each galaxy
         (labeled at right).  The horizontal axis is the SSP age.
         Point size is proportional to the relative weight in the fit
         (normalized to the $V$-band) and the colors and point types
         code the SSP metallicity, as indicated.  The gray shading
         indicates the age range within which the TP-AGB phase is
         active.}
\label{fig:SFHs}
\end{center}
\end{figure}

Figure~\ref{fig:fits} shows the full population synthesis fits to
the central spectra of NGC 628 and UGC 2124 (extracted as described in
Section~\ref{sec:extract}), and in Figure~\ref{fig:SFHs} we plot the inferred
SFHs from the fits.  In Table~\ref{tab:percent}, we list for each
galaxy the percent-light and mass contributions of ``very young''
(0.001--0.4\,Gyr), ``young'' (1--2\,Gyr), ``intermediate''
(4--7\,Gyr), and ``old'' (10--20\,Gyr) age SSPs to the fits.

%\begin{deluxetable}{l|cccc|cccc@{}}
\begin{deluxetable*}{lcccccccc}
\tabletypesize{\footnotesize}
\tablewidth{0pt}
\tablecaption{Percent Contributions of all SSPs in Given Age Ranges 
(Very Young\,=\,0.001--0.4\,Gyr; Young\,=\,1--2\,Gyr;
Intermediate\,=\,4--7\,Gyr; Old\,=\,10--20\,Gyr) to Full Population
Synthesis Fits Weighted by Light ($V$-band normalized) and
Mass. \label{tab:percent} }
\tablehead{
\multicolumn{1}{l}{} &
\multicolumn{4}{c}{Light Weight} &
\multicolumn{4}{c}{Mass Weight} \\
\cline{2-5}\cline{6-9}\\
\multicolumn{1}{l}{Name} &
\multicolumn{1}{c}{0.001--0.4} &
\multicolumn{1}{c}{1--2} &
\multicolumn{1}{c}{4--7} &
\multicolumn{1}{c}{10--20} &
\multicolumn{1}{c}{0.001--0.4} &
\multicolumn{1}{c}{1--2} &
\multicolumn{1}{c}{4--7} &
\multicolumn{1}{c}{10--20} \\
%\multicolumn{1}{c@{}}{10--20} \\
\multicolumn{1}{l}{} &
\multicolumn{1}{c}{(Gyr)} &
\multicolumn{1}{c}{(Gyr)} &
\multicolumn{1}{c}{(Gyr)} &
\multicolumn{1}{c}{(Gyr)} &
\multicolumn{1}{c}{(Gyr)} &
\multicolumn{1}{c}{(Gyr)} &
\multicolumn{1}{c}{(Gyr)} &
\multicolumn{1}{c}{(Gyr)} 
%\multicolumn{1}{c@{}}{(Gyr)} 
}
\startdata
 N0628 & 7 & 67 &  0 & 26 & 1 & 25 &  0 & 74 \\
 U2124 & 0 & 33 &  0 & 67 & 0 &  7 &  0 & 93
\enddata
\end{deluxetable*}
For both galaxies there is a contribution from 13 to 20\,Gyr
SSPs\footnote{We are not concerned here with model ages that are older
than the age of the universe as absolute model SSP ages are not
precisely calibrated.}.  While their weight in light may not seem
dominant, these old populations contribute 74\% and 93\% to the stellar
mass budget of NGC 628 and UGC 2124, respectively
(Table~\ref{tab:percent}).  However, we are particularly interested
here in a significant contribution from SSPs that would be expected by
the Mar05 models to have NIR excesses from their treatment of the
TP-AGB stellar evolution phase.  The age range for which the signature
is present is highlighted by the gray shading in Figure~\ref{fig:SFHs}
and referred to as ``young'' in Table~\ref{tab:percent}.  The
contribution to the $V$-band normalized light of the young population
of interest is 67\% and 33\% for NGC 628 and UGC 2124, respectively.
Thus, if our SFHs are reliable, according to the Mar05 models, these
galaxies would be expected to show signs of the TP-AGB NIR
excess\footnote{Note that we did not perform the population synthesis
fits with the Mar05 models as they have a lower resolution than BC03,
thus reducing the predictive power of the spectra to disentangle the
high frequency features in the fitting process.  A direct comparison
between fits using different models awaits the release of higher
resolution models.}.

\section{``Predicted'' Versus Observed Optical--NIR Colors }
\label{sec:predicted}

\begin{figure}
\begin{center}
\includegraphics[width=0.49\textwidth]{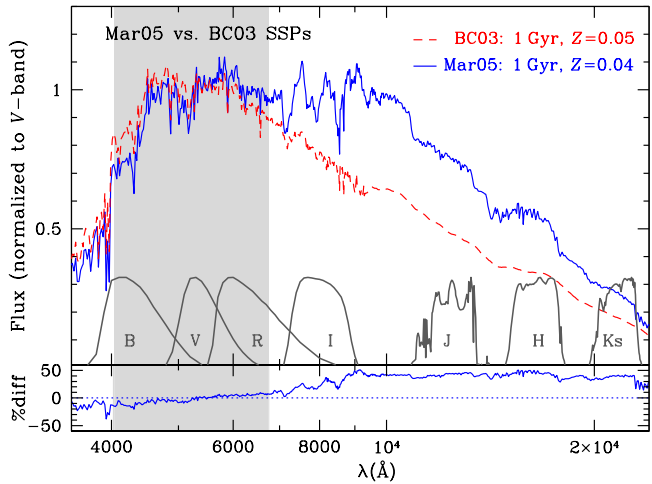}
%\plotone{f3.eps}
\caption{Comparison of SSP SEDs for the Mar05 and BC03 models.  The
         dashed red line is the 1\,Gyr/$Z$\,=\,0.05 SSP from the BC03
         models.  The solid blue line is the 1\,Gyr/$Z$\,=\,0.04 SSP
         from the Mar05 models.  The relevant filter response curves
         for our observed colors are shown as dark gray curves in the
         upper panel.  The lower panel plots the model differences,
         where \%diff\,=\,100$*$(BC03$-$Mar05)/[(BC03$+$Mar05)/2].
         The gray shading represents the optical range of our GMOS
         spectra within which the population fits are constrained.}
\label{fig:SEDs}
\end{center}
\end{figure}

The full population synthesis fits to the spectra are constrained by
the observed 4050--6750\,\AA\ region (gray shading in
Figure~\ref{fig:SEDs}) and, modulo emission lines, represent this
region very well (see Figure~\ref{fig:fits}).  Since the BC03 models cover
the range 90\,\AA\,--\,160\,$\mu$m, we can use the model fits to
``predict'' colors in other bands.  Any difference between colors
predicted in this fashion and the observed colors can be explained by
any of the following scenarios: unrealistic SFHs due to degeneracies
in the age/$Z$/dust plane that cannot be constrained with optical data
alone; inadequacy of the template library to represent the full
coverage of SP age and $Z$ in any given integrated spectrum;
unrealistic estimate of the dust extinction from the population
synthesis fits; errors in the SP modeling predictions outside the
optical range due to, for example, errors in the stellar evolutionary
physics.

To test for any discrepancies, we now compare the predicted
optical--NIR colors from the full population synthesis BC03 model fits
to the observed colors.  A direct estimate of the predicted NIR excess
is complicated by a number of issues related to differences between
the BC03 and Mar05 models.  First, for a given age and $Z$, the model
predictions are not identical, even within the optical limits of our
spectra.  Second, the SEDs provided in each model do not cover
precisely the same age/$Z$ grid.  To illustrate differences between the
BC03 and Mar05 model SEDs, we compare in Figure~\ref{fig:SEDs} the
1\,Gyr/$Z$\,=\,0.05 SSP from BC03, which contributes significantly
to both SFHs in Figure~\ref{fig:SFHs}, to the SED closest in age and
$Z$ from the Mar05 models (1\,Gyr/$Z$=0.04).  While most of the
difference is attributed to the different treatments of the TP-AGB
phase between BC03 and Mar05, some of it could also be due to the
overall larger contribution of AGB stars in the Mar05 models.  For
this comparison, the BC03 models were smoothed with a boxcar of
FWHM\,=\,20\,\AA\ and resampled to $\Delta\lambda$\,=\,20\,\AA/pix
to roughly match the resolution and sampling of the Mar05 SEDs.

\begin{figure}
\begin{center}
\includegraphics[width=0.49\textwidth,bb=18 330 592 718]{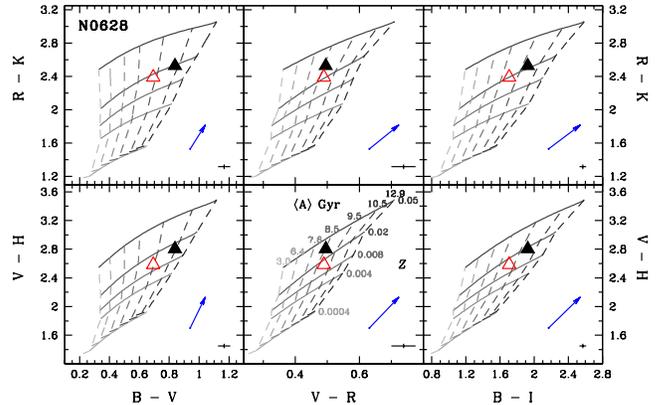} 
%\plotone{f4.eps}
\caption{Color-color plots for NGC 628.  The red open triangles are
	 colors measured (``predicted'') from the BC03 model full
	 population synthesis fits to the observed central spectrum.
	 The black solid triangles are the corresponding observed
	 colors.  Observational errors (calibration and sky
	 subtraction) are indicated at lower right.  Overplotted are
	 grids of BC03 models with exponential SFHs (see MacArthur
	 \etal\ 2004 for details).  Solid and dashed lines are iso-$Z$
	 and iso-age, respectively, with darker shading indicating
	 higher $Z$s and older ages; see middle bottom panels for labels of
	 average age and $Z$.  {\bf Blue arrows:} 
	 model prediction differences between the
	 1\,Gyr/$Z$\,=\,0.05 BC03 SSP (red dashed curve in
	 Figure~\ref{fig:SEDs}) and the Mar05 1\,Gyr/$Z$\,=\,0.04 SSP
	 (solid blue curve in Figure~\ref{fig:SEDs}).  The arrows point toward
	 the Mar05 predictions (from their respective
         location for the BC03 models).}
\label{fig:colorcolor}
\end{center}
\end{figure}

\begin{figure}
\begin{center}
\includegraphics[width=0.49\textwidth,bb=18 330 592 718]{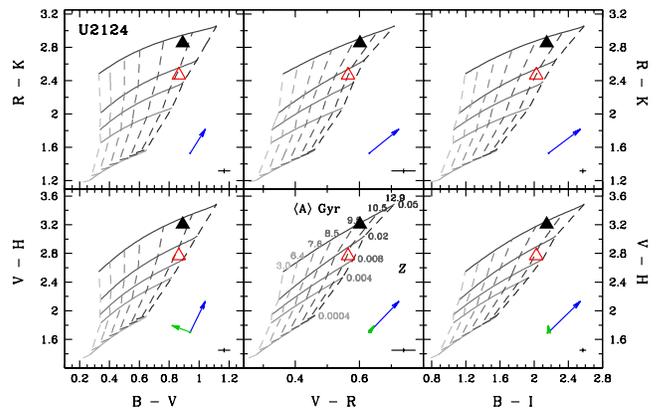} 
%\plotone{f5.eps}
\caption{Same as Figure~\ref{fig:colorcolor}, but for UGC 2124 and the
         {\bf green arrows} show
	 the effect of a non-solar abundance ratio SSP on the colors.
	 Specifically, this represents the difference between 
	 \alphaFe\,=\,0.4 and \alphaFe\,=\,0 stars with
	 T$_{eff}$\,=\,4000\,K, [Fe/H]\,=\,solar, log($g$)\,=\,3.0
	 from the models of Coelho \etal\ (2005).}
\label{fig:colorcolor2}
\end{center}
\end{figure}

Given the above caveats, we now look at the observed versus predicted
optical--NIR colors for NGC 628 in Figure~\ref{fig:colorcolor}.  In
all color combinations, the model fits (red open triangles) predict colors
that are too blue compared to the observed (black solid triangles)
optical--NIR colors.  The blue arrows represent the difference in
color between the BC03 1\,Gyr/$Z$\,=\,0.05 SSP and the Mar05
1\,Gyr/$Z$\,=\,0.04 SSP.  Indeed, the blue vectors represent the
data--model difference very well.  For UGC 2124 in
Figure~\ref{fig:colorcolor2}, it appears that a steeper vector would
be required to precisely match the data--model difference.  In Mac09
we demonstrated by a comparison of absorption-line indices that there
is evidence of an enhanced \alphaFe\ SP in the bulge region of UGC
2124 (see Figure~4 in Mac09), which could also affect the observed
colors (only solar-scaled abundance ratio models are considered here).
To assess whether abundance ratios could compensate for the extra
steepness in the data--model difference, we also plot in
Figure~\ref{fig:colorcolor2} a green vector which represents the
effect of a super-solar abundance ratio star at a given
T$_{eff}$/log(g)/$Z$ from the synthetic stellar models of Coelho
\etal\ (2005) (see the figure caption for details).  The Coelho models do
not extend far enough into the NIR for an accurate $K$-band estimate.
A combined effect of both TP-AGB treatment and super-solar \alphaFe\
would be represented by the addition of the two vectors and resulting
in a steepening that more closely accounts for the data/model offsets.

\begin{figure}
\begin{center}
\includegraphics[width=0.45\textwidth,bb=18 420 592 718]{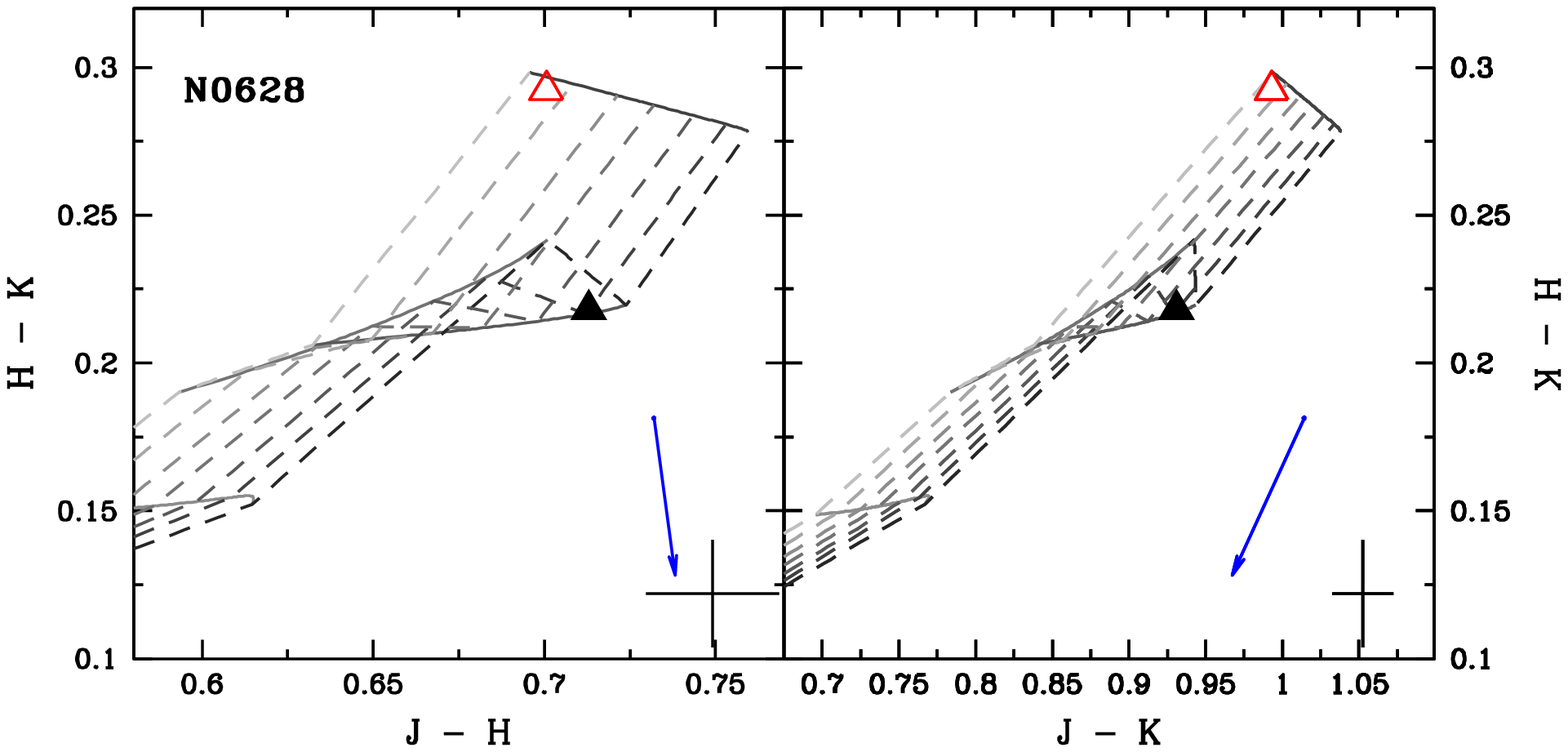} 
\includegraphics[width=0.45\textwidth,bb=18 420 592 718]{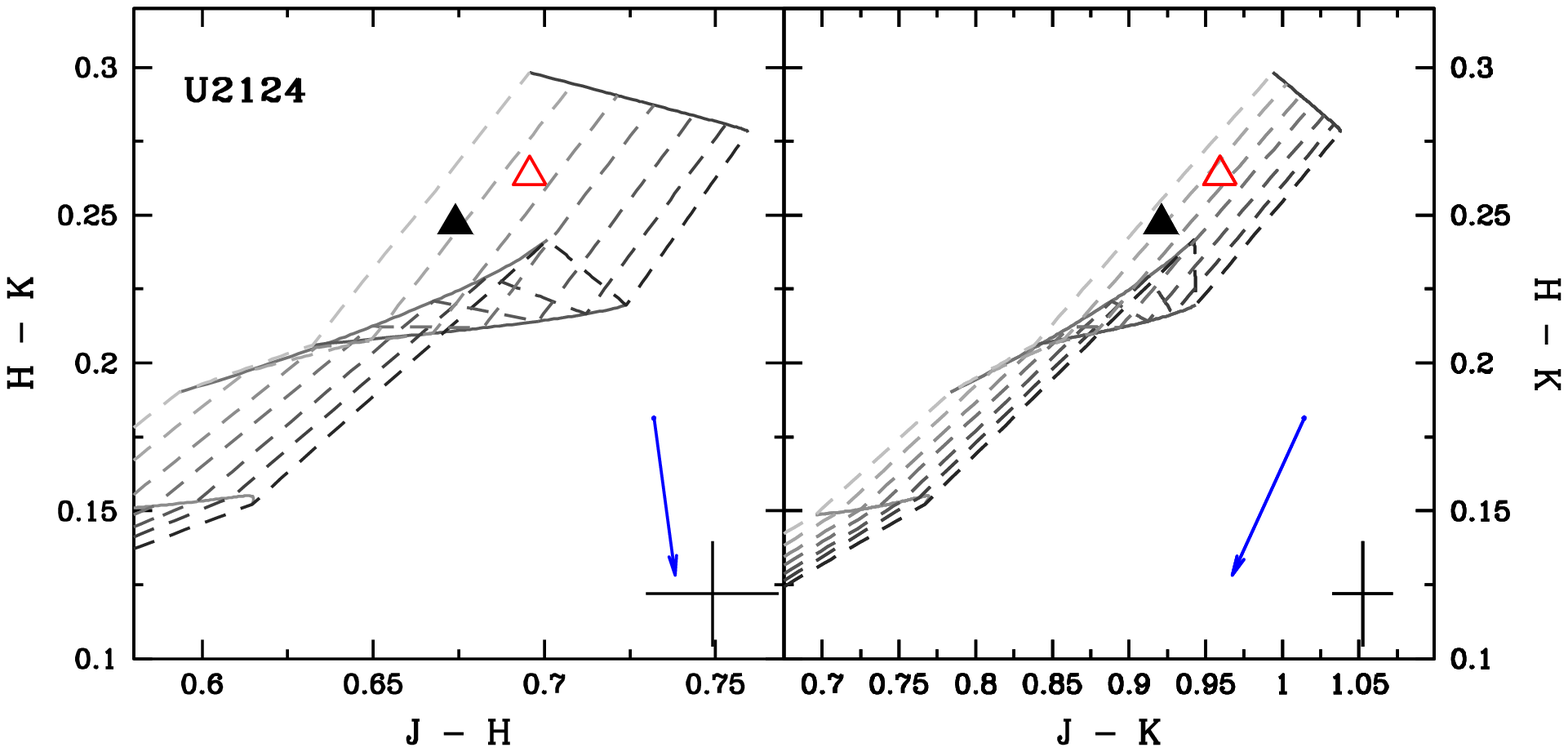} 
%\plotone{f6a.eps}
%\plotone{f6b.eps}
\caption{Same as in Figure~\ref{fig:colorcolor}, but for NIR colors.}
\label{fig:colorcolor2MASS}
\end{center}
\end{figure}

As a further check on the model predictions, we plot in
Figure~\ref{fig:colorcolor2MASS} NIR--NIR colors for both galaxies.
Again we see a difference between the observations and model
predictions, but not in the same sense as the optical--NIR colors.
While the $J-H$ color is predicted to be redder in the Mar05 models,
the $J-K$ and $H-K$ colors are predicted to be bluer.  As for the
optical--NIR colors, the Mar05 models could account for these
differences.  The only mismatch is in the $H-K$ versus $J-H$ colors for
UGC 2124, but this could very well be another manifestation of 
\alphaFe-enhanced SPs (whose predictions do not extend far enough to
test this).

\section{Discussion}\label{sec:discuss}

The above analysis has important implications for the modeling and
interpretations of integrated stellar populations.  The full
population synthesis fitting of NGC 628 and UGC 2124 indicates the
presence of a young component ($\sim$\,1\,Gyr, where Mar05 and BC03
model predictions differ significantly) contributing $\sim$\,50\% to
the $V$-band flux.  In SPS models which account for a detailed
treatment of all stages of the TP-AGB stellar evolution phase, this
young population would contribute a significant amount of NIR flux,
resulting in very red optical--NIR colors, a mild reddening in $J-H$,
and a bluing of the $H-K$ and $J-K$ colors.  All of these predicted
trends were confirmed with the observed colors, thus providing further
support for the presence of the young component, in addition to
strongly favoring the contribution of the TP-AGB phase used in the
Mar05 SPS models (as well as the Marigo \& Girardi (2007) models which
agree well with the Mar05 predictions).

Such anomalous colors have been reported before.  For a sample of 5800
galaxies from the SDSS, Eminian \etal\ (2008) compared optical and NIR
colors with quantities derived from spectra (star formation rate
(SFR), age, $Z$, and dust attenuation).  They found that galaxies with
higher SFRs, while having bluer optical colors ($g-r$), also had {\it
redder} NIR colors ($H-K$), which they interpreted as being in
qualitative agreement with the dominance in NIR light of the TP-AGB
phase after a burst of SF.

At higher redshift, Ellis, Abraham, \& Dickinson (2001) observed the
colors of bulges out to $z\sim$\,1 and found that, while many of the
bulges had blue optical colors, consistent with recent SF, at
$z$\,$\gtrsim$\,0.5, the same bulges had red observed $J-H$ colors.
At these higher redshifts, this translates roughly into rest-frame
$i-J$.  Indeed, while the Mar05 models only show a mild reddening
compared to those of BC03 of the $J-H$ color (by $\sim$\,0.1\,mag for
a solar 1\,Gyr SSP), the $i-J$ color is much redder (by
$\sim$\,0.7\,mag), thus explaining why the effect is only observed at
higher-$z$.  While Ellis \etal\ attributed the red NIR colors to a
predominantly ``short burst'' mode of bulge building at high-$z$, the
red NIR colors could also be explained by the presence of young SPs in
the active TP-AGB phase.

These results indicate that caution must be taken when interpreting the
optical--NIR colors of integrated stellar populations.  Red colors
have typically been interpreted as being due to some combination of
old age, high $Z$, and significant dust extinction.  However, we have
shown here that some of the redness could actually be due to a
significant contribution to the SED from a young SP with significant
NIR flux excess from the TP-AGB evolutionary phase.  Further
confirmation of this result and its implications awaits 
larger samples and analysis with forthcoming higher-resolution
implementations of the SP models which include proper treatment of
the TP-AGB stellar evolutionary phase.

\acknowledgements
We thank Claudia Maraston for stimulating discussions.  
LAM acknowledges financial support from the National
Science and Engineering Council of Canada (NSERC).  SC acknowledges
financial support through a Discover Grant from the NSERC.  This study
is based on
observations obtained at the Gemini Observatory, which is operated by
the Association of Universities for Research in Astronomy, Inc., under
a cooperative agreement with the NSF on behalf of the Gemini
partnership: the National Science Foundation (United States), the
Science and Technology Facilities Council (United Kingdom), the
National Research Council (Canada), CONICYT (Chile), the Australian
Research Council (Australia), Minist{\' r}io da Ci{\^ e}ncia e
Tecnologia (Brazil), and Ministerio de Ciencia, Tecnolog{\' i}a e
Innovaci{\' o}n Productiva (Argentina).


\begin{thebibliography}{}
\bibitem[Bruzual \& Charlot(2003)]{BC03} Bruzual, A.~G. \& Charlot, S.\ 2003,
 \mnras, 344, 1000 (BC03)
% Old and Young Bulges in Late-Type Disk Galaxies
\bibitem[Carollo et al.(2007)]{2007ApJ...658..960C} Carollo, C.~M., 
Scarlata, C., Stiavelli, M., Wyse, R.~F.~G., 
\& Mayer, L.\ 2007, \apj, 658, 960 
\bibitem[Chabrier(2003)]{2003PASP..115..763C} Chabrier, G.\ 2003, \pasp,
 115, 763 
\bibitem[Charlot \& Fall(2000)]{CF00} Charlot, S.~\& Fall, S.~M.\ 2000, \apj,
 539, 718 (C=F00)
\bibitem[Cid Fernandes et al.(2005)]{2005MNRAS.358..363C} Cid Fernandes, R.,
 Mateus, A., Sodr{\'e}, L., Stasi{\'n}ska, G., \& Gomes,
 J.~M.\ 2005, \mnras, 358, 363 
% Library of synthetic stellar spectra with non-solar [alpha/Fe] ratios
\bibitem[Coelho et al.(2005)]{2005A&A...443..735C} Coelho, P., Barbuy, B.,
 Mel{\'e}ndez, J., Schiavon, R.~P., \& Castilho, B.~V.\ 2005, \aap, 443, 735 
\bibitem[Coelho et al.(2007)]{2007MNRAS.382..498C} Coelho, P., Bruzual, G.,
 Charlot, S., Weiss, A., Barbuy, B., \& Ferguson, J.~W.\ 2007, \mnras,
 382, 498 
% Data reduction procedures
\bibitem[Courteau(1996)]{1996ApJS..103..363C} Courteau, S.\ 1996, \apjs, 
103, 363
\bibitem[Ellis et al.(2001)]{2001ApJ...551..111E} Ellis, R.~S., Abraham,
 R.~G., \& Dickinson, M.\ 2001, \apj, 551, 111 
% NIR colours of low-z SDSS galaxies
\bibitem[Eminian et al.(2008)]{2008MNRAS.384..930E} Eminian, C.,
 Kauffmann, G., Charlot, S., Wild, V., Bruzual, G., Rettura, A., \&
 Loveday, J.\ 2008, \mnras, 384, 930 
% MOPED
\bibitem[Heavens et al.(2000)]{2000MNRAS.317..965H} Heavens, A.~F.,
 Jimenez, R., \& Lahav, O.\ 2000, \mnras, 317, 965
% GMOS
\bibitem[Hook et al.(2004)]{2004PASP..116..425H} Hook, I.~M., J{\o}rgensen,
 I., Allington-Smith, J.~R., Davies, R.~L., Metcalfe, N., Murowinski, R.~G.,
 \& Crampton, D.\ 2004, \pasp, 116, 425 [GMOS]
%Standard star reference
\bibitem[Landolt(1992)]{1992AJ....104..340L} Landolt, A.~U.\ 1992, \aj, 
104, 340 
%PEGASE-HR
\bibitem[Le Borgne et al.(2004)]{2004A&A...425..881L} Le Borgne, D.,
 Rocca-Volmerange, B., Prugniel, P., Lan{\c c}on, A., Fioc, M., \&
 Soubiran, C.\ 2004, \aap, 425, 881
%Stellar Population Models and Individual Element Abundances. II.
% Stellar Spectra and Integrated Light Models
\bibitem[Lee et al.(2009)]{2009ApJ...694..902L} Lee, H.-c., et al.\ 2009,
 \apj, 694, 902 
% Structure of Disk-dominated Galaxies. I. Bulge/Disk Parameters, 
% Simulations, and Secular Evolution
\bibitem[MacArthur \etal\ (2004)]{Mac04} MacArthur, L.~A., Courteau, S.,
 Bell, E., \& Holtzman, J.~A. \ 2004, \apjs, 152, 175
\bibitem[MacArthur et al.(2003)]{Mac03} MacArthur, L.~A., 
 Courteau, S., \& Holtzman, J.~A.\ 2003, \apj, 582, 689
\bibitem[MacArthur et al.(2009)]{Mac09} MacArthur, L.~A., Gonz{\' a}lez,
 J.~J., \& Courteau, S.\ 2009, MNRAS, 395, 28 [Mac09]
% TP-AGB in models
\bibitem[Maraston(2005)]{2005MNRAS.362..799M} Maraston, C.\ 2005, \mnras, 
362, 799  [Mar05]
% TP-AGB models for high-z galaxies
\bibitem[Maraston et al.(2006)]{2006ApJ...652...85M} Maraston, C., Daddi, 
E., Renzini, A., Cimatti, A., Dickinson, M., Papovich, C., Pasquali, A., 
\& Pirzkal, N.\ 2006, \apj, 652, 85 
\bibitem[Marigo \& Girardi(2007)]{2007A&A...469..239M} Marigo, P., \&
 Girardi, L.\ 2007, \aap, 469, 239 
% red central B-I vs I-H bulges colours
\bibitem[Peletier et al.(1999)]{1999MNRAS.310..703P} Peletier, R.~F., 
Balcells, M., Davies, R.~L., Andredakis, Y., Vazdekis, A., Burkert, A., 
\& Prada, F.\ 1999, \mnras, 310, 703 
\bibitem[Schiavon(2007)]{2007ApJS..171..146S} Schiavon, R.~P.\ 2007, \apjs,
 171, 146 
\bibitem[Schlegel et al.(1998)]{1998ApJ...500..525S} Schlegel, D.~J.,
 Finkbeiner, D.~P., \& Davis, M.\ 1998, \apj, 500, 525 
% 2MASS reference
\bibitem[Skrutskie et al.(2006)]{2006AJ....131.1163S} Skrutskie, M.~F.,
 et al.\ 2006, \aj, 131, 1163 [2MASS] 
\bibitem[Walcher et al.(2006)]{2006ApJ...649..692W} Walcher, C.~J.,
 B{\"o}ker, T., Charlot, S., Ho, L.~C., Rix, H.-W., Rossa, J., Shields,
 J.~C., \& van der Marel, R.~P.\ 2006, \apj, 649, 692 
\end{thebibliography}
\end{document}